\documentstyle[12pt,psfig]{article}
\newcommand{\eV}{~\mbox{eV}}

\newcommand{\GeV}{~\mbox{GeV}}
\newcommand{\TeV}{~\mbox{TeV}}
\newcommand{\gsim}{ \mathop{}_{\textstyle \sim}^{\textstyle >} }
\newcommand{\lsim}{ \mathop{}_{\textstyle \sim}^{\textstyle <} }
\newcommand{\vev}[1]{ \left\langle {#1} \right\rangle }
\begin{document}%%%%%%%%%%%%%
%%%%%%%%%%%%%%%%%%%%%%%%%%%%%
%%%%%%%%%%%%%%%%%%%%%%%%%%%%%
%%%%%%%%%%%%%%%%%%%%%%%%%%%%%

\baselineskip 1.33em

\setcounter{footnote}{0}
 
%%%%%%%%%%%%%%%%%%%%%%%%%%%%%%%%%%%%%%%%%%%%%%%%%%%%%%%%
\begin{titlepage}%%%%%%%%%%%%%%%%%%%%%%%%%%%%%%%%%%%%%%%
%%%%%%%%%%%%%%%%%%%%%%%%%%%%%%%%%%%%%%%%%%%%%%%%%%%%%%%%
\begin{flushright}
UT-02-14\footnote{UT-999}\\
hep-ph/0203189\\
\end{flushright}

\vskip 1cm
\begin{center}
 {\large \bf Predictions on the neutrinoless double beta decay\\
 from the leptogenesis via the $LH_{u}$ flat direction}
 
 \vskip 1.2cm 

 Masaaki Fujii$^1$,  K. Hamaguchi$^1$, and T. Yanagida$^{1,2}$
 \vskip 0.4cm

 {\it $^1$ Department of Physics, University of Tokyo, Tokyo 113-0033,
 Japan}\\
 {\it $^2$ Research Center for the Early Universe, University of Tokyo,
 Tokyo, 113-0033, Japan}

\vskip 2cm

\abstract{If the baryon asymmetry in the present universe is generated
 by decays of the $L H_u$ flat direction, the observed baryon asymmetry
 requires the mass of the lightest neutrino to be much smaller than the
 mass scale indicated from the atmospheric and solar neutrino
 oscillations. Such a small mass of the lightest neutrino leads to a
 high predictability on the rate of the neutrinoless double beta
 ($0\nu\beta\beta$) decay. In this letter we show general predictions on
 the $0\nu\beta\beta$ decay in the leptogenesis via the $LH_u$ flat
 direction.  }

\end{center}
%%%%%%%%%%%%%%%%%%%%%%%%%%%%%%%%%%%%%%%%%%%%%%%%%%%%%%%%
\end{titlepage}%%%%%%%%%%%%%%%%%%%%%%%%%%%%%%%%%%%%%%%%%
%%%%%%%%%%%%%%%%%%%%%%%%%%%%%%%%%%%%%%%%%%%%%%%%%%%%%%%%

\setcounter{footnote}{0}

~

\section{Introduction}

Leptogenesis~\cite{FY} has been widely considered as an attractive
mechanism to account for the baryon asymmetry in the present universe,
especially after the Super-Kamiokande Collaboration reported the
evidence of the atmospheric neutrino oscillation~\cite{SK-Atm}. In
particular, the leptogenesis via the $L H_u$ flat direction~\cite{MY},
based on the Affleck-Dine mechanism~\cite{AD}, seems to work naturally
in the supersymmetric (SUSY) standard model. Recently, we have performed
detailed analyses on this leptogenesis mechanism via the $L H_u$ flat
direction, taking into account the relevant thermal
effects~\cite{AFHY,FHY},\footnote{The thermal effects can be avoided if
we assume a gauged U$(1)_{B-L}$~\cite{FHY2} or a low scale
inflation~\cite{Asaka}.} and have shown that the generated
baryon asymmetry is determined almost only by the mass of the lightest
neutrino $m_\nu$~\cite{FHY}. In order to explain the empirical baryon
asymmetry in the present universe, the mass of the lightest neutrino
should be much smaller than the mass scale of the atmospheric and solar
neutrino oscillations. This fact leads to a high predictability on the
rate of the neutrinoless double beta ($0\nu\beta\beta$)
decay~\cite{FHY}.

In the previous work, we have assumed the normal hierarchy in the
neutrino mass spectrum. In this letter, we investigate predictions on
the $0\nu\beta\beta$ decay in the more general cases including the
inverted hierarchy in neutrino masses. As we will see, it is quite
interesting that the predicted value of the mass parameter $m_{\nu_{e}
\nu_{e}}$ is in the accessible range of future
experiments~\cite{GENIUS,CUORE,MOON,XMASS,EXO,Majorana,bb-Review} of the
$0\nu\beta\beta$ decay.

\section{Leptogenesis via the $L H_u$ flat direction}

In this section we briefly review the leptogenesis via the $L H_u$ flat
direction~\cite{MY,AFHY,FHY,MM}. Let us start by writing down the
effective dimension-five operator in the superpotential,
%%%
\begin{eqnarray}
 \label{EQ-LHu-start}
 W = 
  \frac{1}{2 M_i}
  \left(L_i H_u \right)
  \left(L_i H_u \right)
  \,,
\end{eqnarray}
which induces small neutrino masses $m_{\nu_{i}}$~\cite{seesaw} after
the neutral component of the Higgs field $H_u$ obtains its vacuum
expectation value $\vev{H_u} = 174\GeV\times \sin\beta$,\footnote{ ${\rm
tan}\beta$ is defined as ${\rm tan}\beta\equiv \vev{H_{u}}/\vev{H_{d}}$,
where $H_u$ and $H_{d}$ are the Higgs fields which provide masses for
the up- and down-type quarks, respectively.

}
\begin{equation}
m_{\nu_i}=\frac{\vev{H_{u}}^2}{M_{i}},
\end{equation}
where we have taken a basis in which the neutrino
mass matrix is diagonal. We adopt the following supersymmetric D-flat
direction~\cite{MY}:
%%%
\begin{eqnarray}
 \widetilde{L} = 
  \frac{1}{\sqrt{2}}
  \left(
   \begin{array}{c}
    \phi\\
    0
   \end{array}
   \right)
   \,,
   \quad
   H_u = 
   \frac{1}{\sqrt{2}}
   \left(
    \begin{array}{c}
     0\\
     \phi
    \end{array}
    \right)
    \,.
\end{eqnarray}
%%%
Here and hereafter, we suppress the family index $i$. As we will see
below, the flat direction which generates the lepton asymmetry most
effectively corresponds to the {\it lightest} neutrino.

The total scalar potential for the flat direction field
$\phi$ is given by
%%%
\begin{eqnarray}
 V(\phi) &=&
  m_{\phi}^2 |\phi|^2 
  + \frac{m_{3/2}}{8 M} \left(a_m \phi^4 + H.c.\right)
  \nonumber\\ &&
  -\,\, c_H H^2|\phi|^2
  + \frac{H}{8 M} \left(a_H \phi^4 + H.c.\right)
  \nonumber\\ &&
  +\sum_{f_k|\phi| < T} c_k f_k^2 T^2  |\phi|^2
  + a_g \alpha_S(T)^2 \,T^4\ln\left(\frac{|\phi|^2}{T^2}\right)
  \nonumber\\ &&
  +\,\, \frac{1}{4 M^2}|\phi|^6
  \,.
\end{eqnarray}
%%%
Here, the potential terms in the first line comes from the SUSY breaking
at the zero temperature $T = 0$, which are mediated by supergravity, and
we take $m_{\phi}\simeq m_{3/2}|a_m|\simeq 1\TeV$,
hereafter.\footnote{We assume gravity mediated SUSY breaking models.} 
The terms in the second line, depending on the Hubble parameter $H$,
reflects the additional SUSY breaking effects caused by the finite
energy density of the inflaton~\cite{DRT}. Hereafter, we take $c_H\simeq
1(>0)$ and $|a_H|\simeq 1$. The terms in the third line represent the
effects of the finite temperature~\cite{ACE,AFHY,SomeIssues,FHY}. Here,
$f_k$ represent Yukawa and gauge coupling constants of the field $\phi$,
$\alpha_S$ is the gauge coupling constant of the SU$(3)_C$, and $c_k$
and $a_g$ are coefficients of order unity.\footnote{The list of $c_k$
and $f_k$ is given in Ref.~\cite{AFHY}. In the case of $L H_u$ flat
direction, $a_g$ is given by $a_g = 9/8$.} Finally, the last term
directly comes from the superpotential in Eq.~(\ref{EQ-LHu-start}).

During the inflation, the negative Hubble mass term $-c_H H^2|\phi|^2$
causes an instability of the flat direction field $\phi$ around the
origin ($\phi\simeq 0$), and $\phi$ field runs to one of the following
four minima of the potential:
\begin{eqnarray} 
 |\phi| &\simeq& \sqrt{M H}
  \,,
  \label{EQ-LHu-minima}
  \\
 \arg(\phi) &\simeq& \frac{- \arg(a_H) + ( 2 n + 1 )\pi}{4}
  \,,
  \qquad n = 0\cdots 3
  \,,
\end{eqnarray}
which are the balance points between the Hubble-induced terms and the
$|\phi|^6$ term.

After the inflation, the amplitude of the $\phi$ field starts to be
reduced following the gradual decrease of the Hubble parameter $H$ [see
Eq.~(\ref{EQ-LHu-minima})], and eventually, either thermal terms $T^2
|\phi|^2$ or $T^4 \ln (|\phi|^2)$, or the soft mass term
$m_{\phi}^2|\phi|^2$ dominate the scalar potential, which causes a
coherent oscillation of the $\phi$ around the origin. At this stage, the
difference between the phases of the complex couplings $a_m$ and $a_H$
leads to a phase rotational motion of $\phi$, which results in
generation of the lepton asymmetry, $n_L = (1/2)i(\dot{\phi}^* \phi -
\phi^* \dot{\phi})$. Decay of $\phi$ produces the lepton asymmetry in
the thermal bath and a part of the produced lepton asymmetry is then
converted into the baryon asymmetry~\cite{FY} via the sphaleron
effects~\cite{KRS}.

\begin{figure}[t!]%%%%%%%%%%%%%%%%%%%%%%%%%%%%%%%%%%%%%%%%%%%
%%%%%%%%%%%%%%%%%%%%%%%%%%%%%%%%%%%%%%%%%%%%%%%%%%%%%%%%%%%%
 \centerline{\psfig{figure=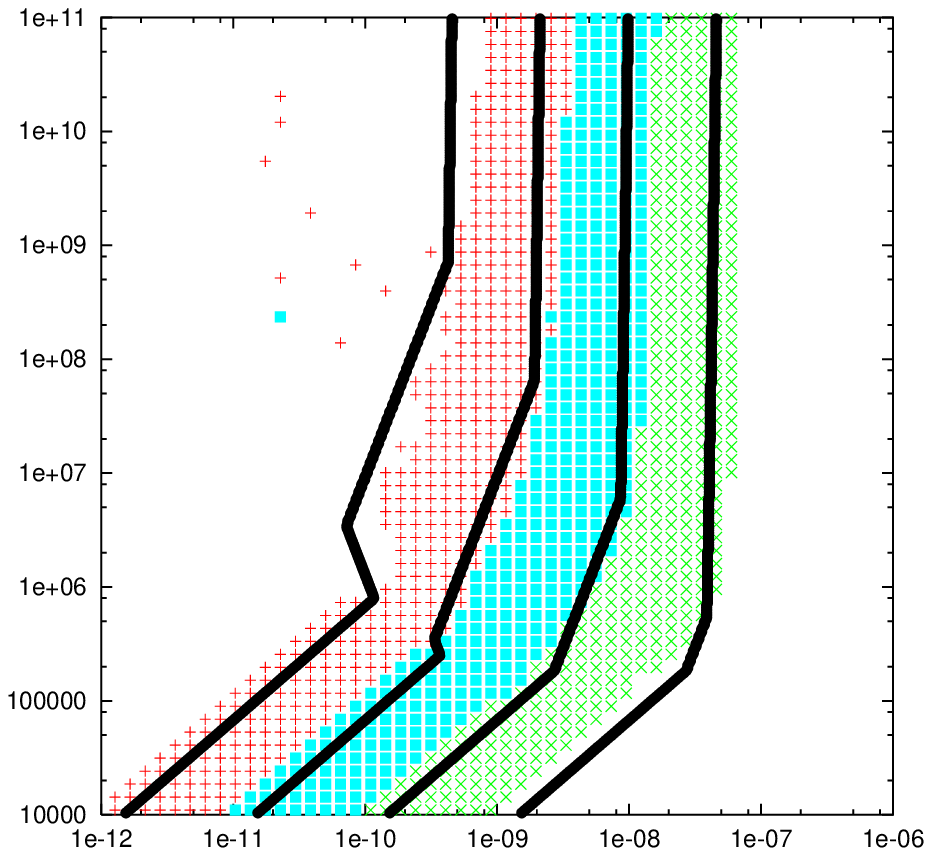,height=10cm}}
 %%%%%%%%%%%%%%%%%%%%%%%%%
 \begin{picture}(0,0)%%%%%
 %%%%%%%%%%%%%%%%%%%%%%%%%
  \put(50,150){$T_R$}  
  \put(43,133){$[$GeV$]$}  
  \put(230,1){$m_\nu\,\,[$eV$]$}
 %%%%%%%%%%%%%%%%%%%%%%%%%
 \end{picture}%%%%%%%%%%%%
 %%%%%%%%%%%%%%%%%%%%%%%%%
 \caption{The plots of baryon asymmetries $n_B/s$ in the $m_\nu$--$T_R$
 plane. The solid lines are the contour plot of the $n_B/s$ obtained by
 analytical calculation, which represent $n_B/s = 10^{-9}$, $10^{-10}$,
 $10^{-11}$, and $10^{-12}$ from the left to the right. The regions with
 points show the result of the numerical simulation, which represent
 $10^{-9} > n_B/s > 10^{-10}$, $10^{-10} > n_B/s > 10^{-11}$ and
 $10^{-11} > n_B/s > 10^{-12}$ from the left to the right. In the
 numerical simulation, we have taken $m_\phi = m_{3/2}|a_m| = 1\TeV$,
 $c_H = |a_H| = 1$, $\arg (a_m) = 0$ and $\arg (a_H) = \pi/3$. } 
 \label{FIG-BAU}
%%%%%%%%%%%%%%%%%%%%%%%%%%%%%%%%%%%%%%%%%%%%%%%%%%%%%%%%%%%%
\end{figure}

We show the result of detailed calculation~\cite{FHY} in
Fig.~\ref{FIG-BAU}, which presents the contour plot of the baryon
asymmetry in the neutrino mass ($m_\nu$) -- reheating temperature
($T_R$) plane. As can be seen in Fig.~\ref{FIG-BAU}, the baryon
asymmetry is almost independent of the reheating temperature $T_R$, and
is determined only by the mass of the lightest neutrino $m_\nu$ for
$T_{R}\gsim 10^5\GeV$. Thus, the baryon asymmetry in the present
universe $n_B/s\simeq (0.4$--$1)\times 10^{-10}$ indicates the mass of
the lightest neutrino to be
\begin{eqnarray}
 m_\nu \simeq (0.1 - 3)\times 10^{-9}\eV\,
\end{eqnarray}
in a wide range of the reheating temperature $10^5\GeV \lsim T_R\lsim
10^{12}\GeV$.\footnote{If the U(1)$_{B-L}$ symmetry is gauged in high
energy scale, the resultant baryon asymmetry is enhanced in the ``D-term
stopping case''~\cite{FHY2}. However, even in this case, the required
mass of the lightest neutrino must satisfy $m_{\nu}\lsim 10^{-5}\eV$ to
avoid the cosmological gravitino problem, and the prediction on the
$0\nu\beta\beta$ decay is almost the same as that presented in this
letter. }

\section{Prediction on the $0\nu\beta\beta$ decay}

The neutrinoless double beta $(0\nu\beta\beta)$ decay, 
if observed, is the strongest evidence 
for lepton number violation.  In other words, it suggests the
Majorana character of the neutrinos and thus the existence of the 
nonrenormalizable operator given in Eq.(\ref{EQ-LHu-start}),
which is a crucial ingredient for our leptogenesis to work.

In the present scenario, as explained in the previous section, the
observed baryon asymmetry in the universe is mainly determined by the
mass of the lightest neutrino, $m_{\nu}\sim 10^{-9}\eV$, regardless of
the reheating temperature of inflation.  This small neutrino mass
predicted from the leptogenesis, combined with the atmospheric and solar
neutrino oscillation experiments, allows us to have definite predictions
on the rate of the $0\nu\beta\beta$ decay.  This is a clear contrast to
the other baryo/leptogenesis scenarios which heavily rely on various
unknown parameters in high energy physics.

In our previous publication~\cite{FHY}, we have assumed the normal mass
hierarchy for the light neutrinos. In this letter, we investigate
general predictions on the $0\nu\beta\beta$ decay taking into account of
the case of inverted mass hierarchy for neutrinos.

The most important ingredient to determine the rate of the
$0\nu\beta\beta$ decay is the effective mass of the electron-type
neutrino, which is given by
\begin{equation}
|m_{\nu_{e}\nu_{e}}|=|U_{e1}^2 m_{\nu_{1}}+U_{e2}^{2}m_{\nu_2}+U_{e3}^{2}
 m_{\nu_3}|,
\label{e-typemass}
\end{equation}
where $m_{\nu_i}$'s are the mass eigenvalues of the neutrinos.
$U_{\alpha i}$ is the mixing matrix which diagonalizes the neutrino mass
matrix, where $\alpha=e,\;\mu,\;\tau$ represent the weak eigenstates.
Here, we take a basis in which the mass matrix of the charged lepton is
diagonal.
%%%%%%%%%%%%%%%%%%%%%%%%%%%%%%%%%%%%%%%%%%%%%%%%%%%%%%%%%%%%
\begin{figure}[t!]%%%%%%%%%%%%%%%%%%%%%%%%%%%%%%%%%%%%%%%%%%%
%%%%%%%%%%%%%%%%%%%%%%%%%%%%%%%%%%%%%%%%%%%%%%%%%%%%%%%%%%%%
 \centerline{\psfig{figure=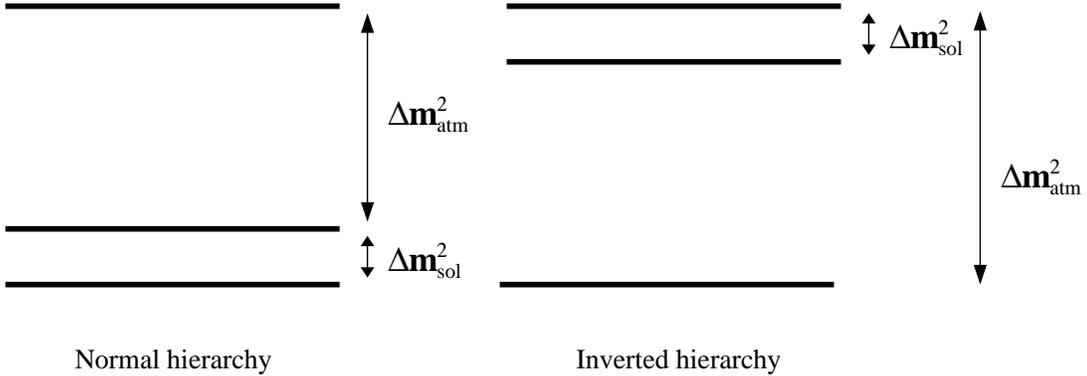,height=5cm}}
 \caption{Hierarchies in the neutrino mass spectrum.}
\label{FIG-masspattern}
%%%%%%%%%%%%%%%%%%%%%%%%%%%%%%%%%%%%%%%%%%%%%%%%%%%%%%%%%%%%
\end{figure}%%%%%%%%%%%%%%%%%%%%%%%%%%%%%%%%%%%%%%%%%%%%%%%%
%%%%%%%%%%%%%%%%%%%%%%%%%%%%%%%%%%%%%%%%%%%%%%%%%%%%%%%%%%%%

In general, the mass pattern for the neutrinos can be classified into
the two cases; the normal mass hierarchy and the inverted mass
hierarchy, which are presented in Fig.~\ref{FIG-masspattern}.  In this
letter, we label the each mass eigenstates in the following way; $
m_{\nu_1}<m_{\nu_2}<m_{\nu_3} $ for the case of the normal mass
hierarchy, and $ m_{\nu_3}<m_{\nu_1}<m_{\nu_2}\; $ for the case of the
inverted mass hierarchy.  Then, the observed mass squared differences are
given by
\begin{equation}
\Delta m^{2}_{\rm atm}=|m_{\nu_3}^{2}-m_{\nu_2}^{2}|,\qquad
\Delta m^{2}_{\rm sol}=m_{\nu_2}^{2}-m_{\nu_1}^{2}\;,
\end{equation}
for the atmospheric and the solar neutrino oscillations, respectively.
By taking this convention, we can treat the mixing angles of the
neutrinos in a similar way for both cases of the neutrino mass
hierarchies.  The mixing matrix $U_{\alpha i}$ is parameterized as
\begin{eqnarray}
U_{\alpha i}=\left(
\begin{array}{ccc}
c_{12}c_{13}& s_{12}c_{13}& s_{13}e^{-i \delta}\\
-s_{12}c_{23}-c_{12}s_{23}s_{13}e^{i\delta}&
c_{12}c_{23}-s_{12}s_{23}s_{13}e^{i\delta}& s_{23}c_{13}\\
s_{12}s_{23}-c_{12}c_{23}s_{13}e^{i\delta}&
-c_{12}s_{23}-s_{12}c_{23}s_{13}e^{i\delta}& c_{23}c_{13}
\end{array}
\right)\cdot\;P\;,
\end{eqnarray}
where $c_{ij}\equiv {\rm cos}\theta_{ij}$, $s_{ij}\equiv {\rm
sin}\theta_{ij}$ and $P={\rm diag}(1,e^{i\beta},e^{i\gamma})$.  $\delta$
is a Dirac-type phase and $\beta, \gamma$ are two phases associated with
the Majorana character of the neutrinos.  By virtue of our convention,
the $\theta_{23}$ and $\theta_{12}$ always correspond to the mixing
angles for the atmospheric and the solar neutrino oscillations,
respectively, regardless of the type of mass hierarchy for the
neutrinos.  In addition, the element of the mixing matrix $U$
constrained by the CHOOZ experiment is always correspond to $(e,3)$
element as $|U_{e3}|\lsim 0.15$~\cite{CHOOZ}.

The requirement for our leptogenesis 
to explain the observed baryon asymmetry
is that the mass of the
lightest neutrino is to be $m_{\nu}\sim 10^{-9}\eV$, which is negligibly 
small compared with the other two mass eigenvalues of the neutrinos.
This immediately leads to the following upper and lower bounds 
on the electron-type neutrino mass:
\begin{eqnarray}
&&\left||U_{e2}|^2 m_{\nu_2}-|U_{e3}|^{2}m_{\nu_3}\right|
\leq|m_{\nu_e\nu_e}|\leq|U_{e2}|^{2}m_{\nu_2}+|U_{e3}|^{2}m_{\nu_3}\;,
\label{mnene-Normal}\\
&&
\left|
|U_{e1}|^{2}m_{\nu_1}-|U_{e2}|^{2}m_{\nu_2}
\right|\leq |m_{\nu_e\nu_e}|\leq
|U_{e1}|^{2}m_{\nu_1}+|U_{e2}|^{2}m_{\nu_2}
\label{mnene-Invert}
\end{eqnarray}
for the normal and the inverted mass hierarchies, respectively.  

\begin{figure}[t!]%%%%%%%%%%%%%%%%%%%%%%%%%%%%%%%%%%%%%%%%%%%
%%%%%%%%%%%%%%%%%%%%%%%%%%%%%%%%%%%%%%%%%%%%%%%%%%%%%%%%%%%%
 \centerline{\psfig{figure=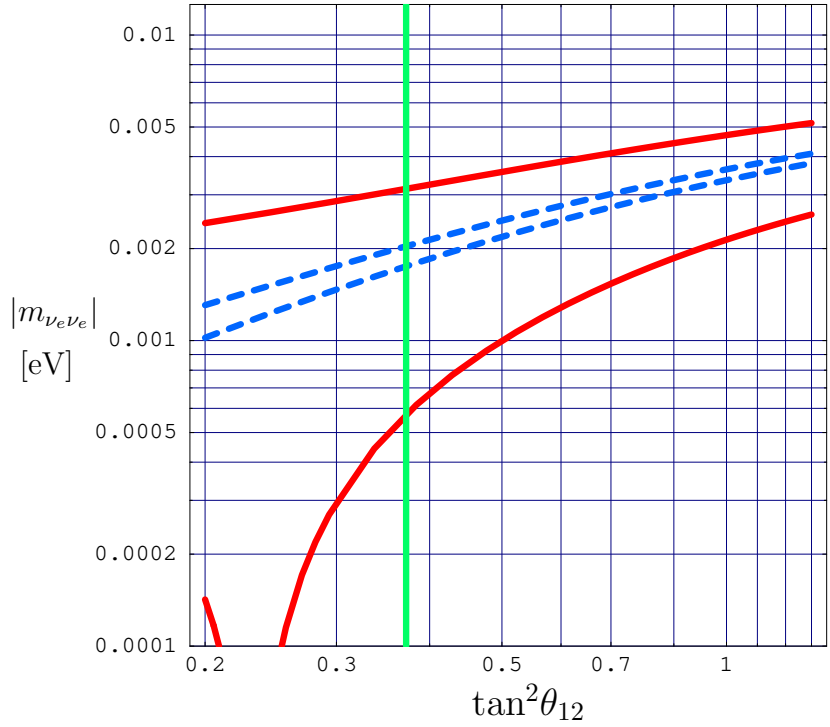,height=9cm}}
 %%%%%%%%%%%%%%%%%%%%%%%%%
 \begin{picture}(0,0)%%%%%
 %%%%%%%%%%%%%%%%%%%%%%%%%
  \put(55,150){$|m_{\nu_e\nu_e}|$}  
  \put(59,130){$[$eV$]$} 
\large  \put(230,1){${\rm tan}^2{\theta_{12}}$}
\normalsize
 %%%%%%%%%%%%%%%%%%%%%%%%%
 \end{picture}%%%%%%%%%%%%
 %%%%%%%%%%%%%%%%%%%%%%%%%
 \caption{The upper and lower bands on the effective mass of the
 electron-type neutrino $|m_{\nu_e\nu_e}|$ in the case of the normal
 mass hierarchy. The red (solid) and blue (dashed)
 lines correspond to the cases where $|U_{e3}|=0.15$ and
 $|U_{e3}|=0.05$, respectively. As for the mass squared differences, we
 have adopted the best fit values. 
The green (vertical) line corresponds 
 to the best fit values of ${\rm tan}^{2}\theta_{12}$ for the LMA solution.}  \label{FIG-Normal}
%%%%%%%%%%%%%%%%%%%%%%%%%%%%%%%%%%%%%%%%%%%%%%%%%%%%%%%%%%%%
\end{figure}

The predictions on the $0\nu\beta\beta$ decay in the case of the normal
mass hierarchy have been investigated in detail in
Ref.~\cite{FHY}.
The upper and lower bounds on the electron-type neutrino mass in
Eq.~(\ref{mnene-Normal}) can be written as
\begin{equation}
|m_{\nu_e\nu_e}|^{\rm max}_{\rm min}
\simeq s_{12}^2 \sqrt{\Delta m^{2}_{\rm sol}}\pm |U_{e3}|^{2}
\sqrt{\Delta m^{2}_{\rm atm}}\;.
\label{Normal-bounds}
\end{equation}
In Fig.~\ref{FIG-Normal}, we present the numerical result 
for the case of the large mixing angle (LMA)
solution. The red (solid) and blue (dashed)
lines correspond to the cases where $|U_{e3}|=0.15$ and $|U_{e3}|=0.05$,
respectively. 
As for the mass squared differences, we have adopted the 
following best fit values~\cite{SK-Atm,Lisi}:
\begin{equation}
\Delta m^{2}_{\rm atm}=3.2\times 10^{-3}\eV^2\;,\qquad
\Delta m^{2}_{\rm sol}=4.9 \times 10^{-5}\eV^{2}\;.
\label{sol-LMA}
\end{equation}
The green (vertical) line denotes the best fit values of the mixing
angle ${\rm tan}^{2}\theta_{12}=0.37$.  The behavior of the bounds when
we vary the mass squared differences $\Delta m^{2}_{\rm atm}$ and
$\Delta m^{2}_{\rm sol}$ is easily seen from the
Eq.~(\ref{Normal-bounds}). We see that the overall scale of the
$|m_{\nu_{e}\nu_{e}}|$ is almost proportional to the $\sqrt{\Delta
m^{2}_{\rm sol}}$. For example, when we vary the $\Delta m^{2}_{\rm
sol}$ within the $95\%$ C.L. allowed region of the LMA solution,
$|m_{\nu_e\nu_e}|$ changes within about $\times/\div 1.7$ of the value
presented in Fig.~\ref{FIG-Normal}.  As clearly shown in this figure,
the $|m_{\nu_e\nu_e}|$ is predicted in a narrow range when the
$|U_{e3}|$ is much smaller than the present bound.
% values by the long base line experiments in the future, such as
%JHF~\cite{JHF}.

Now, let us consider the case of the inverted mass hierarchy for the 
neutrinos. The Eq.~(\ref{mnene-Invert}) can be written 
in terms of the observables in neutrino oscillation experiments as
\begin{eqnarray}
&&|m_{\nu_e\nu_e}|_{\rm max}\simeq (1-|U_{e3}|^{2})\sqrt{\Delta m^{2}_{\rm atm}}
\left(1-\frac{\Delta m^{2}_{\rm sol}}{2 \Delta m^{2}_{\rm atm}} c_{12}^{2}
\right)\;,\label{mnene-max}\\
&&|m_{\nu_e\nu_e}|_{\rm min}\simeq 
(1-|U_{e3}|^{2})\sqrt{\Delta m^{2}_{\rm atm}}\;c_{12}^{2}\left|\left(
(1-{\rm tan}^{2}\theta_{12})-\frac{\Delta m^{2}_{\rm sol}}{2 \Delta 
m^{2}_{\rm atm}}
\right)\right|\;, \label{mnene-min}
\end{eqnarray}
for the upper and lower bounds, respectively.  As seen from these
relations, the mass squared differences for the solar neutrino
oscillations $\Delta m^{2}_{\rm sol}$ gives negligible  effects 
on the upper bound
of $|m_{\nu_e\nu_e}|$. The $\Delta m^{2}_{\rm sol}$ affects the lower
bound of the $|m_{\nu_e\nu_e}|$ only when the following relation is 
satisfied:
\begin{equation}
|(1-{\rm tan}^{2}\theta_{12})|\lsim \frac{\Delta m^{2}_{\rm sol}}{\Delta m^{2}_{\rm atm}}\;.
\label{cond-deltamsol}
\end{equation}

\begin{figure}[t!]%%%%%%%%%%%%%%%%%%%%%%%%%%%%%%%%%%%%%%%%%%%
%%%%%%%%%%%%%%%%%%%%%%%%%%%%%%%%%%%%%%%%%%%%%%%%%%%%%%%%%%%%
 \centerline{\psfig{figure=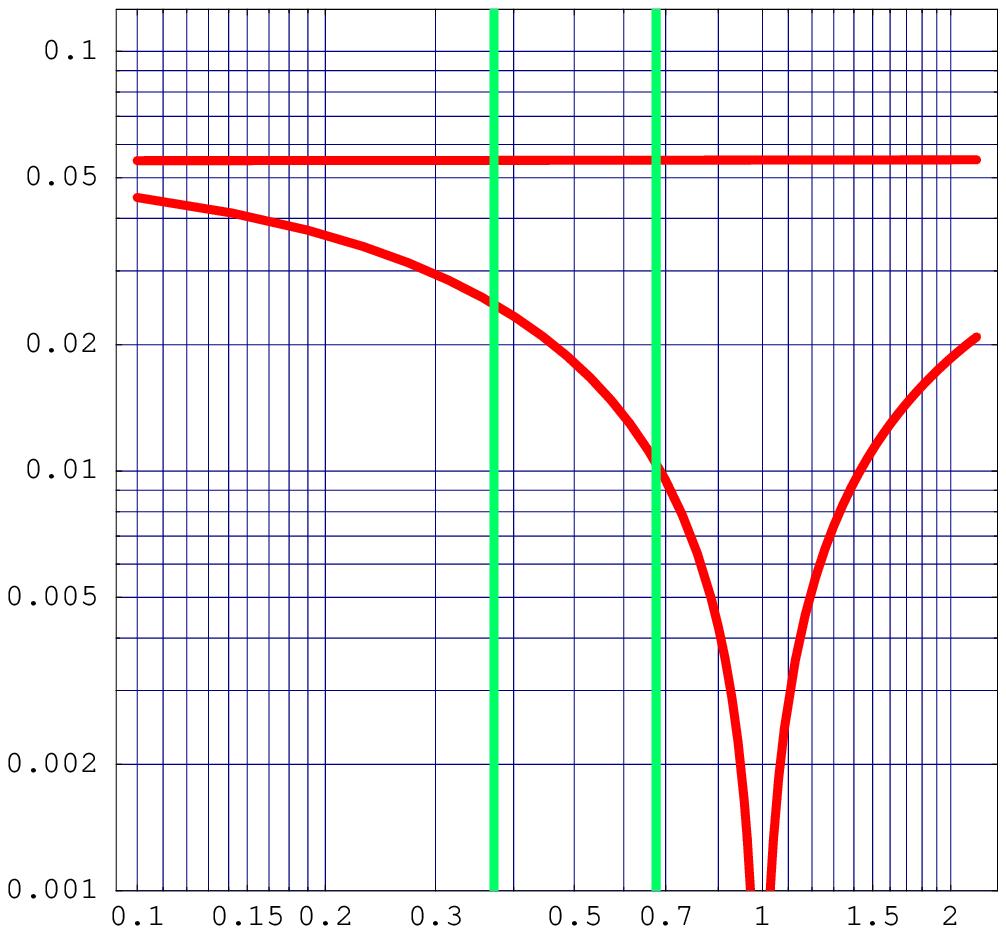,height=9cm}}
 %%%%%%%%%%%%%%%%%%%%%%%%%
 \begin{picture}(0,0)%%%%%
 %%%%%%%%%%%%%%%%%%%%%%%%%
  \put(55,150){$|m_{\nu_e\nu_e}|$}  
  \put(59,130){$[$eV$]$} 
\large  \put(230,1){${\rm tan}^2{\theta_{12}}$}
\normalsize
 %%%%%%%%%%%%%%%%%%%%%%%%%
 \end{picture}%%%%%%%%%%%%
 %%%%%%%%%%%%%%%%%%%%%%%%%
 \caption{The upper and lower bands on the effective mass of the
 electron-type neutrino $|m_{\nu_e\nu_e}|$ in the case of the inverted
 mass hierarchy. Here, we have used the
 $|U_{e3}|=0.15$. As for the mass squared differences, we
 have adopted the best fit values for the LMA solution
 given in Eq.~(\ref{sol-LMA}).
However, this result is applicable to the other solutions for 
the solar neutrino oscillations except for
${\rm tan}^{2}\theta_{12}\simeq 1$ as explained in the text. 
The two green (vertical) lines correspond to 
the best fit values of the mixing angles for the LMA and the LOW
 solutions from left to right, respectively.
 } \label{FIG-Inverted}
%%%%%%%%%%%%%%%%%%%%%%%%%%%%%%%%%%%%%%%%%%%%%%%%%%%%%%%%%%%%
\end{figure}

In Fig.~\ref{FIG-Inverted}, we present the upper and the lower bounds 
on the $|m_{\nu_e\nu_e}|$ in the case of the inverted mass hierarchy.
Here, we have used the $|U_{e3}|=0.15$. As for the mass squared
differences, we have adopted the best fit values for the LMA
solution as before. However, this result is 
applicable to the other solutions for the solar neutrino oscillations
except for the case of ${\rm tan}^{2}\theta_{12}\simeq1$, because of the 
reason mentioned above. 
The two green (vertical) lines correspond to the best fit values of
the mixing angles for the LMA and the LOW solutions $({\rm tan}^{2}
\theta_{12}=0.37,\;0.68)$~\cite{Lisi} from left to right,
respectively.
As seen from the figure, 
the $|m_{\nu_e\nu_e}|$ is restricted in a very small range
such as $0.01\eV\lsim |m_{\nu_e\nu_e}|\lsim 0.06\eV$
when ${\rm tan}^{2}\theta_{12}\lsim 0.7$,
which is clearly in the reach of the future $0\nu\beta\beta$
decay experiments. Even if we change the $\Delta m^{2}_{\rm atm}$ 
within the $99\%$ C.L. allowed region, the overall scale of the 
$|m_{\nu_e\nu_e}|$ varies within only $\times/\div 1.5$ of the value
presented in Fig.~\ref{FIG-Inverted}.

\section{Conclusions}

Leptogenesis via the $LH_{u}$ flat direction is one of the most
interesting scenarios to explain the observed baryon asymmetry, in which
the observed baryon asymmetry has a direct connection to the mass of the
lightest neutrino regardless of the details of high energy physics.  In
this letter, we have derived predictions on the rate of the
$0\nu\beta\beta$ decay, which is regarded as a low energy consequence of
the present leptogenesis; the mass of the lightest neutrino must be
$m_{\nu}\sim 10^{-9}\eV$.

Both in the cases of the normal and the inverted mass hierarchies for
the neutrinos, the region of $|m_{\nu_e\nu_e}|$ is strongly restricted
and it is accessible in the future $0\nu \beta \beta $ decay
experiments~\cite{GENIUS,CUORE,MOON,XMASS,EXO,Majorana,bb-Review}.
Furthermore, predictions on the $|m_{\nu_e\nu_e}|$ become much more
definite when the $|U_{e3}|$ is severely constrained (for the case of
the normal hierarchy) or the mixing angle for the solar neutrino
oscillation is restricted as ${\rm tan}^{2}\theta_{12}<1$ (for the case
of the inverted hierarchy) in the future experiments.  The predictions
on the $0\nu\beta\beta$ decay derived in this letter will play a crucial
role on testing the leptogenesis via the $LH_{u}$ flat direction.

%%%%%%%%%%%%%%%%%%%%%%%%%%%%%%%%%%%%%%%%%%%%%%%%%%%%%%%%%%%%%%%%%%%
%%%%%%%%%%%%%%%%%%%%%%%%%%%%%%%%%%%%%%%%%%%%%%
%%%%%%%%%%%%%%%%%%%%%%%%%%%%%%%%%%%%%%%%%%%%%%%%%%%%%%%%%%%%%%%%%%%

\newpage
%%%%%%%%%%%%%%%%%%%%%%%%%%%%%%%%%%%%%%%%%%%%%%%%%%%%%%%%%%%%
%%%%%%%%%%%%%%%%%%%%%%%%%%%%%%%%%%%%%%%%%%%%%%%%
%%%%%%%%%%%%%%%%%%%%%%%%%%%%%%%%%%%%%%%%%%%%%%%%%%%%%%%%%%%%

%%%%%%%%%%%%%%%%%%%%%%%%%%%%%%%%%%%%%%%%%%%%%%%%%%%%%%%%%%%%
%%%%%%%%%%%%%%%%%%%%%%%%%%%%%%%%%%%%%%%%%%%%%%%%
%%%%%%%%%%%%%%%%%%%%%%%%%%%%%%%%%%%%%%%%%%%%%%%%%%%%%%%%%%%%

%%%%%%%%%%%%%%%%%%%%%%%%%%%%%%%%%%%%%%%%%%%%%%%%%%%%%%%%%%%%
%%%%%%%%%%%%%%%%%%%%%%%%%%%%%%%%%%%%%%%%%%%%%%%%
%%%%%%%%%%%%%%%%%%%%%%%%%%%%%%%%%%%%%%%%%%%%%%%%%%%%%%%%%%%%

%%%%%%%%%%%%%%%%%%%%%%%%%%%%%
%%%%%%%%%%%%%%%%%%%%%%%%%%%%%
%%%%%%%%%%%%%%%%%%%%%%%%%%%%%
\end{document}